# Photoelectroanalytical Oxygen Detection with Titanate Nanosheet – Platinum Hybrids Immobilised into a Polymer of Intrinsic Microporosity (PIM-1)

Bingbing Fan,[a, b] Yuanzhu Zhao,[a] Budi Riza Putra,[a, c] Christian Harito,[d, e] Dmitry Bavykin,[e] Frank C. Walsh,[e] Mariolino Carta,[f] Richard Malpass-Evans,[g] Neil B. McKeown,[g] and Frank Marken*[a]

**Abstract:** The polymer of intrinsic microporosity PIM-1 is employed to disperse and deposit a Pt@titanate nanosheet photocatalyst film. The resulting microporous films allow electrolyte and oxygen permeation to give conventional oxygen reduction voltammetric responses on glassy carbon or on platinum disk electrodes in the dark. Preliminary data are presented showing that with pulsed light from a blue LED (385 nm) oxygen reduction at the electrode is effectively "switched off". A mechanism is tentatively assigned as photocatalytic depletion of oxygen near the electrode. Photoelectrochemical current responses are observed in aqueous NaOH, NaCl, $Na_2HPO_4$ and shown to be light intensity and oxygen concentration dependent. Electroanalytical applications are suggested.

## 1 Introduction

Photoelectroanalytical methods are useful as an alternative to conventional electroanalytical methods based only on potential stimulation. A light pulse is employed at constant applied potential to trigger a photocurrent response as a measure of the analyte concentration. This can be beneficial when electronic components for stepping or sweeping the electrode potential are not desirable. Sometimes selectivity towards the target analyte is introduced/improved based on the photoelectrochemical mechanism, producing selectively reactive intermediates. Broadly, two types of photo-electrochemical sensors can be distinguished:

- Light pulses lead to direct charge carrier production (for example electrons and holes in semiconductor films). The target analyte then reacts with one of these charge carriers to produce a net current signal.
- Light pulses are employed to produce highly reactive intermediates in solution close to the electrode surface. The target analyte then reacts to a product. This either produces the electrochemical signal due to secondary electron transfer, or the signal diminishes due to photochemical consumption of the target analyte.

Experimentally, at constant applied bias voltage, the light on/off current responses can be compared and shown to be selective for certain types of analytes. Light induced electrode mechanisms can be useful also when short-lived redox- or photo-excited intermediates can be employed in analysis (e.g. employing singlet oxygen in drug detection [1,2]). Photo-electroanalytical protocols have been suggested for detection of aminophenol [3], for nitrite detection [4], or for carbohydrate level monitoring [5]. Photoelectrochemical methods have been developed also for biosensors [6,7]. A $TiO_2/Bi_2S_3$ based photoelectrochemical immunosensor for prostate-specific antigen has recently been reported [8]. Here, exploratory experiments are reported for a titanate nanosheet material employed

[a] B. Fan, Y. Zhao, B. R. Putra, F. Marken
Department of Chemistry, University of Bath, Claverton Down, BA2 7AY, UK
E-mail: f.marken@bath.ac.uk
[b] B. Fan
School of Material Science and Engineering, Zhengzhou University, Henan 450001, China
[c] B. R. Putra
Department of Chemistry, Faculty of Mathematics and Natural Sciences,
Bogor Agricultural University, Bogor, West Java, Indonesia
[d] C. Harito
Industrial Engineering Department, Faculty of Engineering, Bina Nusantara University,
Jakarta, Indonesia 11480
[e] C. Harito, D. Bavykin, F. C. Walsh
Energy Technology Research Group, Faculty of Engineering and Physical Science,
University of Southampton, SO17 1BJ, Southampton, UK
[f] M. Carta
Department of Chemistry, Swansea University, College of Science, Grove Building,
Singleton Park, Swansea SA2 8PP, UK
[g] R. Malpass-Evans, N. B. McKeown
EaStCHEM, School of Chemistry, University of Edinburgh, Joseph Black Building,
David Brewster Road, Edinburgh, Scotland EH9 3JF, UK

as photocatalyst immobilised into a polymer of intrinsic microporosity PIM-1.

Polymers of intrinsic microporosity (or PIMs [9]) have been developed based on molecularly highly rigid structures (see molecular structure of PIM-1 in Figure 1). Poor packing in the solid state is responsible for microporosity and for some unusual properties of these materials. PIM-1 has been previously employed in gas adsorption and permeation [10], but also in electrochemical studies for oxygen reduction and hydrogen evolution [11], in photoelectrochemical hydrogen production [12], as artificial solid electrolyte interphase (SEI) in batteries [13,14], or as an ion-selective membrane in redox flow cell systems [15]. PIM-1 is soluble in chloroform and therefore readily applied as a film on electrodes with/without embedded catalysts [16]. PIM-1 has a Brunauer-Emmett-Teller (BET) specific surface area of 600–900 $m^2 g^{-1}$ and typically microporous channels with a diameter in the range of 0.4–0.8 nm [17].

Titanate nanosheet materials have been discovered and developed by Sasaki and coworkers [18,19]. Starting with $TiO_2$ a reaction with cesium salts is employed to synthesise an intermediate, which is then exfoliated into individual nanosheets (containing tetrabutylammonium surface species [20]) dispered in a colloidal solution. The nanosheets are generated with typically 200 nm size [21], 1.2–1.3 nm thickness [22], and chemical composition $H_{0.7}Ti_{1.825}\square_{0.175}O_4 \cdot H_2O$) [23]. Applications of these titanate nanosheets have been proposed in batteries [24], in heterogeneous catalysis [25], in photocatalysis [26], and as a filler in polmer blends [27]. Intercalation of redox active species into the inter-lamellar spaces has been reported [28].

Here, a Pt@titanate nano-composite is produced by photodeposition of Pt nanoparticles and embedded into a PIM-1 host film (see Figure 1). In this fashion, the Pt@titanate photocatalyst is not in direct contact with the electrode surface, *i.e.*, there are no direct photocurrents due to charge carrier flow into the electrode. However, photochemical reactions can affect electrode processes and produce photoelectroanalytical signals by affecting solution species in the micropores close to the electrode surface. Figure 1 shows the experimental configuration with PIM-1 and photocatalyst immobilised on the surface of a glassy carbon or platinum disk electrode. An LED light source is deployed to provide light pulses during voltammetry experiments performed with a potentiostatically controlled, three-electrode cell (working electrode, reference electrode, counter electrode).

In this preliminary study of photoelectrochemical reactions in PIM-1 films, titanate nanosheet photocatalysts with photodeposited platinum (Pt@titanate) are investigated. Photocurrents are shown to be "subtractive", in the sense that oxygen reduction currents are lowered in the presence of light. The PIM-1 photocatalyst film is considered to consume oxygen when exposed to light and the electrochemical reduction of oxygen is suppressed in the presence of light. Photocurrents are shown to be correlated to light intensity and to oxygen concentration. A mechanism is tentatively assigned. Although details of the mechanism are not fully resolved, it is suggested that a broader range of applications of oxygen consuming photocatalytic films may be possible in the future. Further characterisation of these photocatalytic electrode materials and their photo-assisted switching is encouraged.

## 2 Experimental

### 2.1 Reagents

Titanate nanosheet material was synthesised as described previously by Sasaki *et al.* [22] and by Harito *et al.* [29,30]. All chemicals were purchased form Sigma-Aldrich or Fisher Scientific and used without further purification.

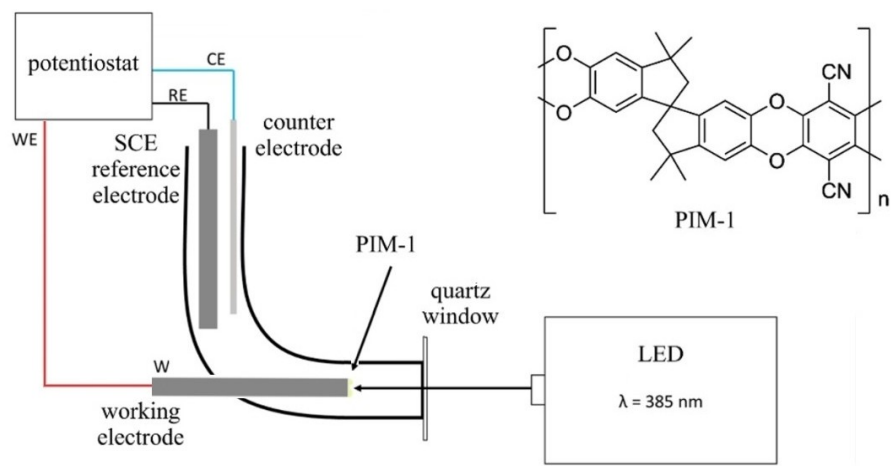

Fig. 1. Molecular structure of PIM-1 and experimental system for photoelectrochemical investigation of oxygen reduction reactions. The distance from the electrode to the LED was approx. $30 \pm 3$ mm and the estimated irradiance at the location of the electrode involved a power density of approx. 100 $mW\,cm^{-2}$.

Aqueous solutions were prepared under ambient condition in volumetric flasks with ultrapure water with resistivity of 18.2 MOhm cm (at 22 °C, taken from an ELGA Purelab Classic system).

### 2.2 Instrumentation

Electrochemical experiments were carried out on an Autolab PGSTAT30 potentiostat (Metrohm, UK) controlling a three-electrode system with a 3 mm diameter platinum or glassy carbon working electrode, a saturated calomel SCE reference electrode and a platinum wire counter electrode. The electrochemical cell had a quartz window, allowing blue LED light ($\lambda=385$ nm, Thorlabs, UK, type M385LP1, operated at 90% nominal power, to give an irradiance of typically 2.1 mW cm$^{-2}$ in 20 cm distance; based on a semiconductor probe measurement at the distance of the electrode roughly 3 cm from the light source this corresponds to a power density of approx. 100 mW cm$^{-2}$) to pass and a $3\pm1$ mm path length of solution to interact with the working electrode (see Figure 1). Transmission electron microscopy (TEM) images were obtained with a JEOL JEM-2100Plus system equipped with an Oxford Instruments X-MaxN TSR Windowless Energy dispersive X-ray analyser (EDX).

### 2.3 Preparation of Photocatalyst

Platinum nanoparticles were photo-attached to the titanate nanosheets by stirring a suspension of 2 mL titanate solution (2.56 mg/mL) and 76 μL K$_2$PtCl$_6$ solution (aqueous 10.3 mM K$_2$PtCl$_6$) with illumination (LED, $\lambda=$ 385 nm, Thorlabs, UK, type M385LP1, operated at 90% nominal power) for at least 72 hours (see Figure 2). The colour changed from white to dark grey. The weight ratio of Pt to titanate in the final product was 2.8 wt% (assuming complete conversion of Pt(VI) to Pt metal). Then the mixture was centrifuged at 8500 rpm for 10 min, the supernatant was separated, and the precipitate was washed 3 times with ethanol, then the final product (Pt@titanate) was dried at 60 °C for 12 h.

The photocatalyst was prepared by mixing 2 mg PIM-1 and 2 mg Pt@titanate powders in 1 mL chloroform, then stirred in solution for at least 72 hours (a long stirring time is required for photo-active products to be obtained). Typically, a volume of 2 μL of the photocatalyst (corresponding to 4 μg PIM-1 and 4 μg Pt@titanate) was applied to the electrode surfaces.

## 3 Results and Discussion

### 3.1 Formation of Pt-loaded Titanate Nanosheets (Pt@titanate)

Titanate nanosheet materials are white in aqueous dispersion (see Figure 3E). In previous work diffraction and TEM data for this material have been published [21]. After addition of K$_2$PtCl$_6$ and exposure to blue light, the colour darkens with time and, after 72 h, a grey product is obtained (see Figure 3E) due to photo-deposition of platinum onto the titanate nanosheet material. The presence of Pt nano-deposits is confirmed by electron micrographs and elemental mapping. As seen in Figure 3, the typical size of Pt nanoparticles on the surface of titanate nanosheets is relatively large (about 50–100 nm). Such poor dispersion is probably due to poor adsorption of [PtCl$_4$]$^{2-}$ anions on the surface of negatively charged titanate nanosheets.

A grey powder was separated by centrifugation. Next, a solution was prepared by dispersing 2 mg PIM-1 and 2 mg Pt@titanate powders in 1 mL chloroform. This solution needed to be stirred for at least 72 h to develop photoactive properties. This could be associated with Pt@titanate dispersing/delaminating only slowly with PIM-1 acting as capping agent for nanosheets in the chloroform phase. A good interaction of Pt@titanate nanosheets and PIM-1 may be important for photo-electrochemical processes to occur. For electrochemical measurements a volume of 2 μL of the colloidal dispersion in chloroform (corresponding to 4 μg PIM-1 and 4 μg Pt@titanate) was deposited onto a 3 mm diameter circular disk electrode.

### 3.2 Photoreduction of Oxygen in a Pt@titanate/PIM-1 Film I.: Photo-Driven Oxygen Removal

A glassy carbon electrode with PIM-1 and Pt@titanate deposit was immersed into aqueous 0.1 M NaOH and the

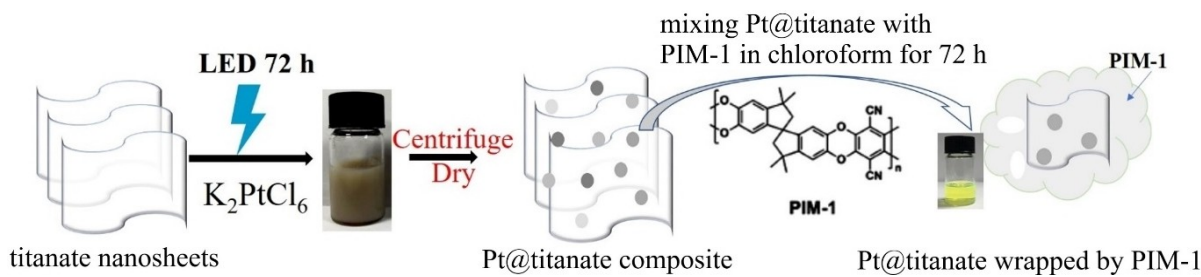

Fig. 2. Schematic illustration of photodeposition of Pt onto titanate nanosheets, purification, and stirring in chloroform with PIM-1 to give delaminated hybrid materials.

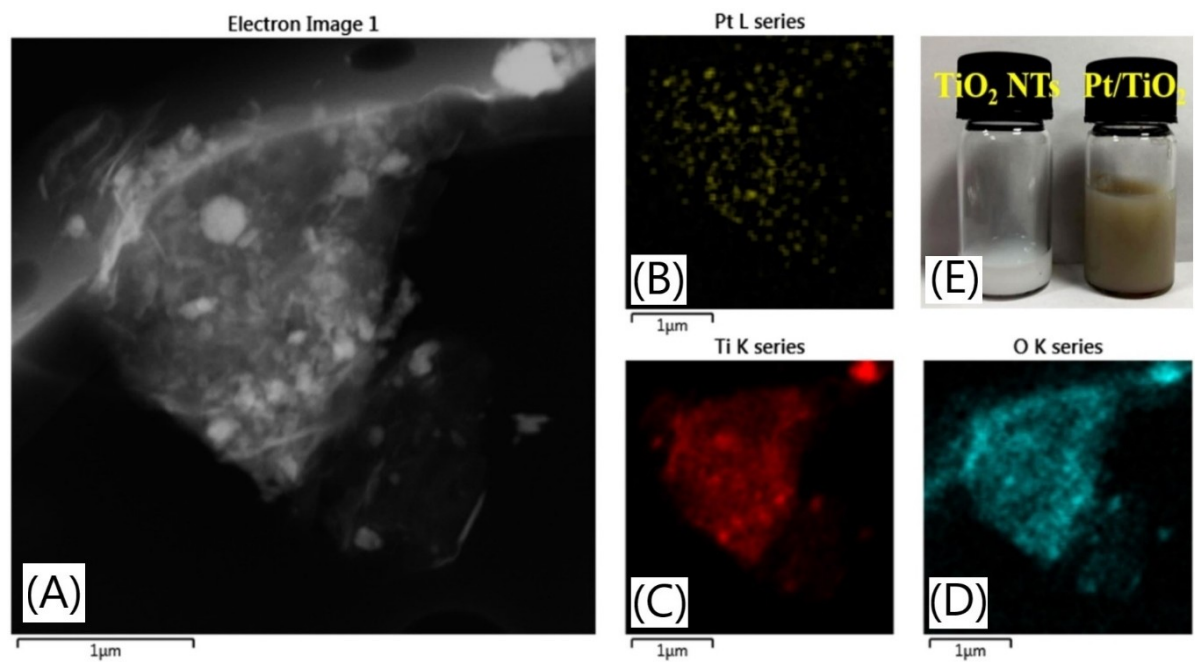

Fig. 3. (A) TEM image of the nanosheets with PIM-1 showing delaminated layered morphology, (B,C,D) EDS elemental analysis of the Pt@titanate nanosheets, and (E) photographs of dispersions of titanate nanosheets with and without Pt deposits.

potential was scanned from 0.0 V vs. SCE negative to −0.8 V vs. SCE and positive to +0.6 V vs. SCE. Figure 4A shows that in the absence of illumination a chemically irreversible reduction peak occurs at −0.52 V vs. SCE. This reduction peak can be identified as oxygen reduction by purging the solution with argon (see Fig-ure 4C). On glassy carbon electrode surfaces the reduction of oxygen is dominated by a 2-electron process producing $H_2O_2$ as an intermediate [31]. On platinum electrodes a 4-electron reduction to water occurs at a less negative potential and with higher peak current (*vide supra*). Data in Figure 4A are obtained with ambient oxygen (oxygen concentration approx. 0.2 mM [32]) and data in Figure 4B are obtained with pure oxygen purged solution (oxygen concentration approx. 1.0 mM).

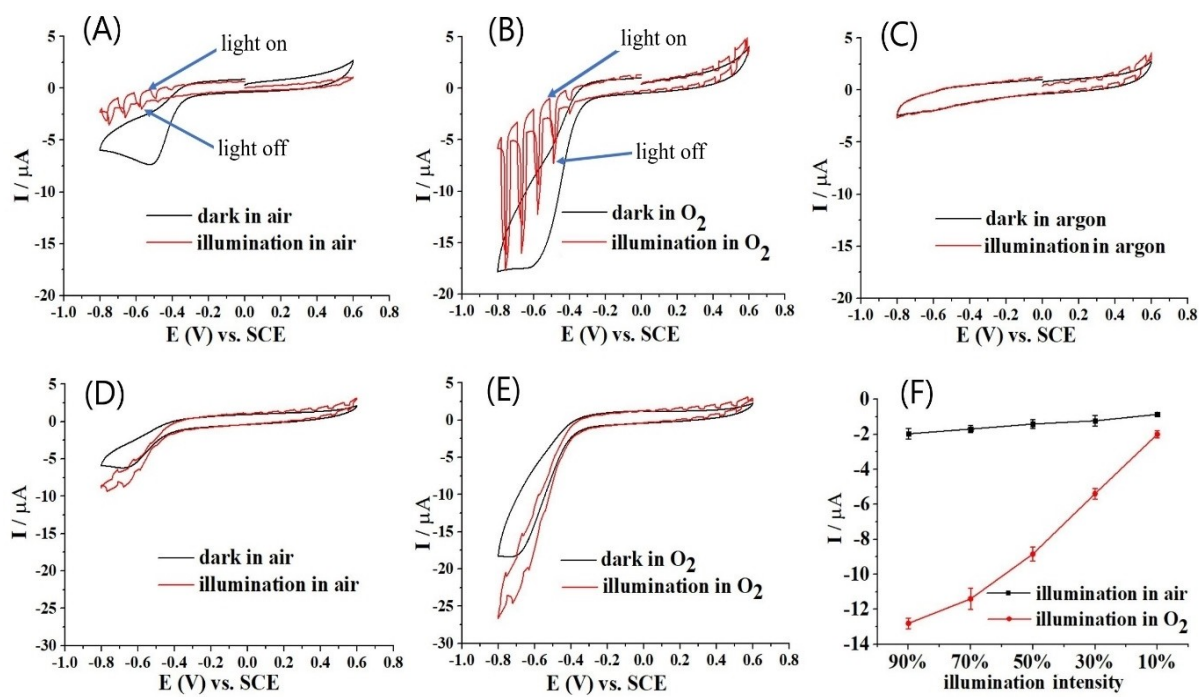

Fig. 4. Cyclic voltammograms (scan rate 30 mV s$^{-1}$; black: no illumination; red: 2 s on and 1 s off, λ=385 nm LED) in 0.1 M NaOH at a glassy carbon electrode (A) coated with Pt@titanate/PIM-1 composite in air-purged solution, (B) as before with pure oxygen purging, (C) as before in solution purged with argon, (D) coated with Pt@titanate without PIM-1 in solution purged with air, (E) as before, in solution purged with pure oxygen. (F) Plot of the photocurrent response *versus* light intensity (conditions as in A,B).

Perhaps interestingly, pulsed light seemed to diminish the reduction current similar to the case when argon purging was applied (see Figure 4A and 4C). When purging the solution with pure oxygen, the oxygen concentration is increased by a factor of 5 and correspondingly the dark reduction current is increased (Figure 4B). However, with pulsed LED light illumination, the current immediately collapses when the light is switched on and the reduction current recovers only when the light is switched off. From the shape of transient photo-current responses it can be observed that the light-on transient is fast (the current quickly decays down to a base level) whereas the light-off transient is somewhat slower and did not fully reach the steady state plateau. The faster oxygen removal is affected by processes within the film; the slower recovery of the signal is likely to be linked to both oxygen diffusion within and outside of the film.

When performing similar experiments without PIM-1 by only depositing Pt@titanate, a very different type of reactivity is observed (Figure 4D,E). In this case, the illumination causes only a slight but insignificant (possibly thermally induced) increase in current. Therefore, the presence of the PIM-1 host material seems important for the oxygen reduction process to occur in the film on the surface of the electrode. Figure 4F shows the clear effect of the light intensity on photocurrents (measured as difference in current during pulses at −0.8 V vs. SCE). A mechanism can be tentatively suggested:

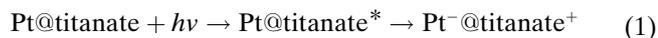 (1)

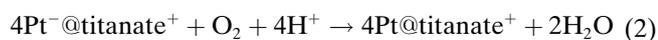 (2)

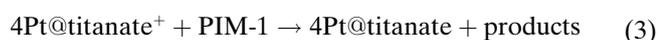 (3)

Assuming that photoexcitation of Pt@titanate triggers charge separation (eqn. 1) and reduction of oxygen occurs in the presence of the Pt nanoparticle (as overall 4-electron reduction; eqn. 2), it may be speculated that the holes are transferred to the organic PIM-1 backbone, which may gradually degrade in this process. This could explain why a long dispersion time of Pt@titanate with PIM-1 was required to improve the interaction of polymer host and photocatalyst. Although plausible, there are currently many unknown factors in this mechanism, such as the additional effect of the PIM-1 (as a fluorescent dye material [33]) on the excitation of the Pt@titanate photocatalyst and reactivity towards oxygen. Also, the long-term effects of illumination on the Pt@titanate/PIM-1 film still need to be investigated.

### 3.3 Photoreduction of Oxygen in a Pt@titanate/PIM-1 Film II: Effect of Electrode Materials

In order to explore the Pt@titanate photoelectrochemical behaviour in more detail, the effects of the electrode material are considered. Figure 5A and 5B compare data for the same experiment in 0.1 M NaOH saturated with pure oxygen with Pt@titanate/PIM-1 deposits for a glassy carbon and for a circular platinum disk electrode (both 3 mm diameter). In the absence of illumination, the reduction of oxygen occurs at −0.5 V vs. SCE on glassy carbon and at −0.3 V vs. SCE on platinum. The higher current during oxygen reduction on platinum is associated with the 4-electron nature (forming water) of the process compared to the reduction on glassy carbon associated with 2-electron reduction (forming $H_2O_2$). With pulsed illumination, a similar effect is observed on both types of electrode consistent with the photocatalytic removal of oxygen close to the electrode surface in eqns (1–3). The nature of the electrode appears to be unimportant and the Pt@titanate/PIM-1 film appears to be similarly effective in both cases.

Data in Figure 5C, 5D, and 5D compare experimental results for a glassy carbon electrode coated with Pt@titanate/PIM-1 and immersed in either 0.1 M NaOH, 0.1 M NaCl, or 0.1 M $NaH_2PO_4$. Very similar behaviour is observed in alkaline solution, neutral solution, and mildly acidic solution, indicating that the photocatalytic oxygen reduction and removal is effective and relatively insensitive to pH or nature of the electrolyte. A possible application of the observed photoelectrochemical reactivity could be oxygen detection.

### 3.4 Photoreduction of Oxygen in a Pt@titanate/PIM-1 Film IV.: Oxygen Sensing in 0.5 M NaCl

In order to explore the possible application of light pulses for oxygen detection in seawater, additional experiments were performed in 0.5 M NaCl (model seawater) and with a gas mixing manifold to mix oxygen and argon. Data in Figure 6A shows the photocurrent responses for a glassy carbon electrode coated with Pt@titanate/PIM-1 in the presence of 100% oxygen. First the effect of the amount of film deposit was evaluated. Data in Figure 6B show the same experiment with twice as much photocatalyst film and data in Figure 6C was obtained with three-fold Pt@titanate/PIM-1 film deposit. A deposition of 8 μg Pt@titanate/PIM-1 appeared to be appropriate with high loadings clearly diminishing the effect (photocurrents don't have time to recover and currents remain low when the light is switched off).

Figure 6D show photocurrent data as a function of oxygen concentration in model seawater solution. Data points have been recorded in triplicate and errors are indicated. With 100% oxygen, the photocurrent ap-

# Full Paper

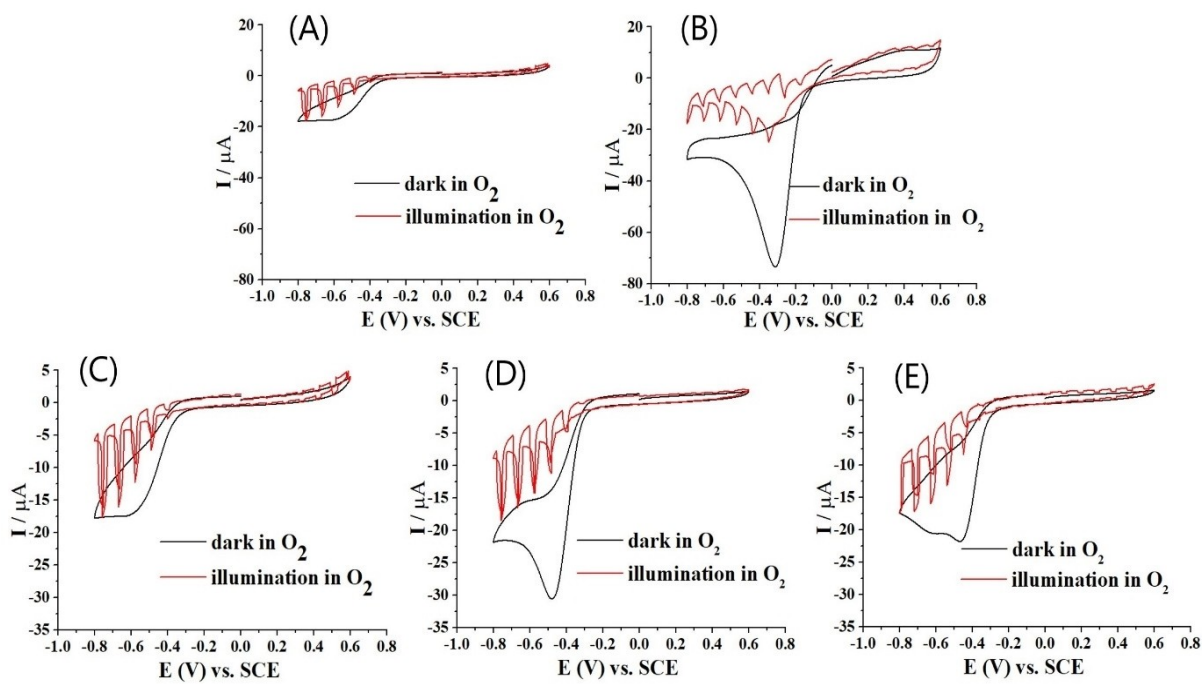

Fig. 5. Cyclic voltammograms (scan rate 30 mV s$^{-1}$; black: no illumination; red: 2 s on and 1 s off, λ=385 nm LED) in 0.1 M NaOH purged with pure oxygen, (A) Pt@titanate/PIM-1 composite at a 3 mm diameter glassy carbon electrode; (B) as before, but at a 3 mm diameter platinum electrode. Cyclic voltammograms (scan rate 30 mV s$^{-1}$; black: no illumination; red: 2 s on and 1 s off, λ=385 nm LED) at a 3 mm diameter glassy carbon electrode coated with Pt@titanate/PIM-1 composites in pure oxygen purged solution containing (C) 0.1 M NaOH, (D) 0.1 M NaCl, (E) 0.1 M NaH$_2$PO$_4$.

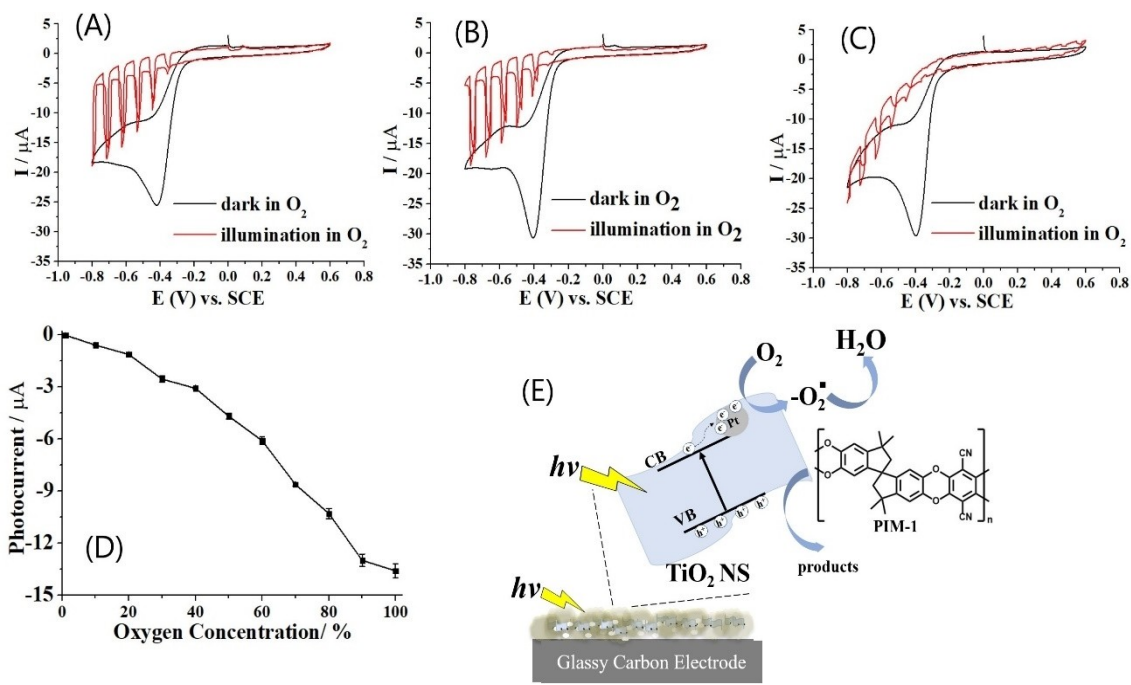

Fig. 6. Cyclic voltammograms (scan rate 30 mV s$^{-1}$; black: no illumination; red: 2 s on and 1 s off, λ=385 nm LED) in 0.5 M NaCl with pure oxygen with different amounts of Pt@titanate/PIM-1composite. (A) 4 μg; (B) 8 μg; (C) 12 μg; (D) Photocurrent responses for oxygen reduction in 0.5 M NaCl, light pulses 2 s on 1 s off, λ=385 nm, applied potential −0.8 V vs. SCE (error bars represent triplicate measurements; two standard deviations). (E) Schematic illustration of the mechanism.

proaches 15 μA at −0.8 V vs. SCE (see Figure 6A) and the current decays at lower oxygen levels. The photocurrent to concentration correlation is not perfectly linear possibly due to kinetic/diffusional factors affecting the sharp light-off component of the signal. Slower measurements with better developed plateau currents may improve linearity of data. But the correlation is clearly observed. Electrodes could be re-used for several hours with good repeatability but more work is required to better explore the long-term behaviour of these electrodes under illumination at different power densities for controlled times.

In the future, the timing of pulses and conditions for the electroanalytical experiment will be optimised to allow more linear correlation of photocurrent to oxygen levels. More work will be required also on interfering species and effects of pressure (water depth) on the analytical response [34]. A key benefit of the photoelectroanalytical approach is that measurements are possible at constant potential without any further electronic components to step or change the applied potential. The light pulse offers an on/off mechanism to periodically subtract the baseline background. However, only further development will reveal whether commercial benefits are sufficiently attractive to encourage real applications.

## 4 Conclusions

Photoelectroanalysis has been demonstrated with a titanate nanosheet catalyst containing photo-deposited platinum. This photocatalyst shows enhanced photocurrents when immobilised into a microporous PIM-1 host material, but only after thorough dispersion. There was no requirement for quencher to be added. Photoelectrochemical hydrogen evolution was not observed, but the reduction of oxygen at the substrate electrode surface was suppressed when the illumination was switched on. It seems likely that photocatalytic depletion of oxygen occurs within the PIM-1 film when exposed to blue LED light. Pulsed light was shown to give oxygen concentration information in 0.5 M NaCl. There are many remaining questions concerning types of photo-processes in this Pt@titanate/PIM-1 hybrid material, the effects introduced by the polymer matrix *versus* effects from the titanate nanosheets, and the role of the Pt nanoparticles aiding $H_2O_2$ intermediate disproportionation as well as aiding the photochemical charge carrier separation. Further work will be required to address these questions.

In the future, photocatalyst films in PIM hosts could be of wider use in photoelectroanalysis and in photocatalysis. PIMs provide a unique microporous environment for photocatalysts to be immobilised and surrounded by the aqueous electrolyte phase in micropores. Diffusion of reactants will be limited to small molecules (such as oxygen) and therefore size selectivity can be imposed. Further effects from the PIM host structure on the photocatalytic reaction and on the electrochemical process need further investigation.
## Acknowledgements

Y.Z. (201809350006) is grateful to the China Scholarship Council for a PhD scholarship. B.F. thanks the China Scholarship Council (CSC scholarship No. 201907045064) for an academic visitor scholarship.

## Data Availability Statement

Additional data are available from f.marken@bath.ac.uk.

# FULL PAPER

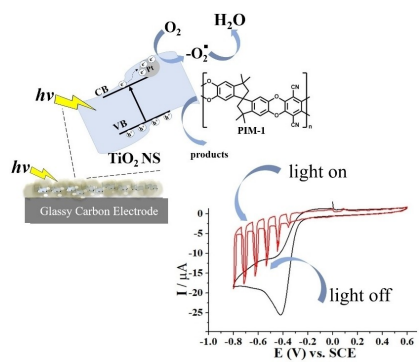

B. Fan, Y. Zhao, B. R. Putra, C. Harito, D. Bavykin, F. C. Walsh, M. Carta, R. Malpass-Evans, N. B. McKeown, F. Marken*

1 – 9

**Photoelectroanalytical Oxygen Detection with Titanate Nanosheet – Platinum Hybrids Immobilised into a Polymer of Intrinsic Microporosity (PIM-1)**